\documentclass[aps,prb,showpacs,twocolumn]{revtex4}
\usepackage{graphicx}
\usepackage{bm}
\begin{document}
\title{Linear-polarization dependence of microwave-induced magnetoresistance oscillations 
in high-mobility two-dimensional systems}
\author{X.~L. Lei and S.~Y. Liu}
\affiliation{Key Laboratory of Artificial Structures and Quantum Control, Department of Physics, Shanghai Jiao Tong University,
800 Dongchuan Road, Shanghai 200240, China}

\begin{abstract}
We examine the effect of changing the linear polarization angle $\theta$ of incident microwaves 
with respect to the dc current on radiation-induced magnetoresistance oscillations in a 
two-dimensional (2D) system within the balance-equation formulation of the photon-assisted 
magnetotransport model, considering the radiative decay as the sole damping mechanism.
At an extremum the amplitude of oscillatory magnetoresistance $R_{xx}$ exhibits a sinusoidal, up to a factor of 5,
 magnitude variation with rotating the polarization angle $\theta$.
The maximal amplitude shows up generally at a nonzero $\theta$, which is dependent upon the extremum in
question, the 2D electron setup, the radiation frequency and the magnetic field orientation.  
These results provide a natural explanation for the experimental observations by Mani {\it et al.} 
[Phys. Rev. B {\bf 84}, 085308 (2011)], and Ramanayaka {\it et al.} [Phys. Rev. B {\bf 85}, 205315 (2012)]. 

\end{abstract}


\maketitle

\section{Introduction}

Microwave induced magnetoresistance oscillation 
in high-mobility two-dimensional (2D) electron systems  
 has been a subject of intensive 
experimental \cite{Zud01,Ye,Mani02,Zud03,Dor03,Yang03,Zud04,Mani04,Willett,Mani05,
Dor05,Stud,Smet05,Yang06,Hatke09-1,Mani10,Mani-PRB11,Mani-PRB12} 
and theoretical \cite{Ryz03,Shi,Durst,Lei03,Koulakov03,Vav04,Dmitriev03,DGHO05,Torres05,
Ng05,Ina-prl05,Kashuba,Mikhailov04,Lei06,Dmitriev09,Ina-PRB07,Ng-PRB08} studies
over the past decades. Though a consensus on the period and phase of the oscillation
has long been reached, the effect of the microwave polarization on the amplitude of the oscillation 
remains one of the most challenging and unsolved issues since the discovery of microwave-induced
magnetoresistance oscillations.   

Early measurement on L-shaped Hall bars indicated that the period
and phase of radiation-induced magnetoresistance oscillations
are insensitive to the relative orientation between the microwave polarization and the current.\cite{Mani05}
A later experiment\cite{Smet05} carried out on specimens with a square geometry in a
quasioptical setup reported a striking result that not only the frequency
and phase but also the amplitude of radiation-induced resistance oscillations and the zero
resistance regions are notably immune to the sense of circular and linear polarizations
of microwaves. This influential result raised a big challenge to the existing theoretical models, in which,
though no detailed investigation was reported, some kind of polarization dependence was believed 
to exist,\cite{Ryz03,Lei03,Koulakov03} 
and thus expedited the emergence of different scenarios 
capable of showing polarization immunity of microwave magnetoresistance response.\cite{Ina-PRB07,Ng-PRB08} 
In a recent study, Mani {\it et al}.\cite{Mani-PRB11} found
a strong sensitivity in the amplitude of radiation-induced
magnetoresistance oscillations to the relative orientation of the
linear polarization with respect to the Hall bar axis. 
Particularly, more detailed measurement\cite{Mani-PRB12} 
by rotating, by an angle $\theta$, the polarization direction of linearly polarized
microwaves with respect to the long axis of the Hall bar electron devices, showed, at relatively low microwave power, 
a strong sinusoidal variation in the diagonal resistance $R_{xx}$ vs $\theta$ at the oscillatory extrema.
And, unexpectedly, the angle for the maximal oscillatory $R_{xx}$ response under a given-power linear-polarized microwave,
which is not at $0^{\rm o}$ or $90^{\rm o}$, appears to depend upon the radiation frequency, the extremum in question, 
and the magnetic field orientation.
So far, there has been no theoretical explanation for these interesting observations. 
There is an urgent need to analyze the detailed polarization dependence from a theoretical
model for radiation-induced magnetoresistance oscillations.

\section{magnetoresistance under polarized radiation}

We deal with an isotropic 2D system of short thermalization time, consisting of $N_{s}$ electrons in a unit area of 
the $x$-$y$ plane. These electrons, scattered by random impurities and by phonons in the lattice, are subjected 
to a uniform magnetic field ${\bm B}=(0,0,B)$ in the $z$ direction.
When an electromagnetic wave of angular frequency $\omega$ illuminates perpendicularly onto the 2D plane 
with the incident electric field 
\begin{equation} 
{\bm E}_{\rm i}(t)={\bm E}_{{\rm i}s}\sin(\omega t)+ {\bm E}_{{\rm i}c}\cos(\omega t) \label{inci}
\end{equation}
at $z=0$ and a dc current flows within the plane, the electric field inside the 2D system
involves a dc component, ${\bm E}_0$, and an ac component
\begin{equation} 
{\bm E}(t)={\bm E}_{s}\sin(\omega t)+ {\bm E}_{c}\cos(\omega t). \label{enc}
\end{equation} 
The steady-transport state of this electron system can be described by the
drift velocity of the electron integrative (the center of mass) motion, consisting of a dc
part, ${\bm v}$, and a stationary time-dependent part, 
\begin{equation}
{\bm v}(t)={\bm v}_s \sin(\omega t)+{\bm v}_c \cos(\omega t),
\end{equation}
in the 2D plane, together with an average temperature $T_{e}$, characterizing the isotropic thermal 
distribution of electrons in the reference frame
moving with the center of mass.\cite{Lei85} They satisfy 
the following force and energy balance equations:\cite{Lei03}
\begin{eqnarray}
&&N_{s}e{\bm E}_{0}+N_{s} e ({\bm  v}\! \times {\bm  B})+
{\bm  F}=0,\label{2deqv0}\\
&& {\bm v}_c=-\frac{e{\bm E}_s}{m \omega}-\frac{{\bm F}_s}{N_{s}m \omega}
-\frac{e}{m \omega}({\bm v}_s\!\times
{\bm B}),\label{2deqv1}\\
&& {\bm v}_s=\,\frac{e{\bm E}_c}{m \omega}+\frac{{\bm F}_c}{N_{s}m \omega}
+\frac{e}{m \omega}({\bm v}_c
\!\times {\bm B}), \label{2deqv2}
\end{eqnarray}
\begin{equation}
 N_{s}e{\bm E}_0\cdot {\bm  v}+S_{\rm  p}- W=0.\label{2deqsw}
\end{equation}
Here,
\begin{equation}
{\bm F}=\sum_{{\bm q}_\|}\left| U({q}_\|)\right| ^{2}
\sum_{n=-\infty }^{\infty }{\bm q}_\|{J}_{n}^{2}(\xi )
{\it \Pi}_2({q}_\|,\omega_0-n\omega )\label{exf0}
\end{equation}
is the time-averaged damping force against the electron drift motion due to impurity scattering, and
\begin{equation}
S_{\rm p}=\sum_{{\bm q}_\|}\left| U({q}_\|)\right| ^{2}
\sum_{n=-\infty }^{\infty } n\omega {J}_{n}^{2}(\xi )
{\it \Pi}_2({q}_\|,\omega_0-n\omega )\label{exsp}
\end{equation} 
is the time-averaged rate of the 
electron energy-absorption from the radiation field.
In Eqs.\,(\ref{exf0}) and (\ref{exsp}), $U({q}_\|)$ is the effective impurity potential,
 ${\it \Pi}_2({q}_\|,{\it \Omega})$ 
is the imaginary part of the electron density-correlation function 
at temperature $T_{e}$ in the presence of the magnetic field without the electric field, 
$\omega_0\equiv{\bm q}_\|\cdot {\bm v}$, and  $J_{n}(\xi )$ 
is the Bessel function of order $n$ with argument 
 $\xi\equiv [{({\bm q}_\|\cdot {\bm v}_s)^2+({\bm q}_\|\cdot {\bm v}_c)^2}]^{\frac{1}{2}}/\omega$.
Note that, although contributions of phonon scattering to ${\bm F}$ and $S_{\rm p}$
are neglected in comparison with those of impurity scattering at the considered low lattice temperature, 
it provides the main channel for electron energy dissipation to the lattice 
with a time-averaged energy-loss rate $W$, having an expression as given in Ref.\,\onlinecite{Lei03}. 

The ac components ${\bm v}_s$ and ${\bm v}_c$ of electron drift velocity should be determined 
selfconsistently in terms of the incident ac field ${\bm E}_{\rm i}(t)$ by the electrodynamic equations 
connecting both sides of the 2D system, taking into account the scattering-related  
damping forces ${\bm F}_s$ and ${\bm F}_c$. \cite{Lei03} However, for high-mobility systems 
at low temperatures, the effects of these scattering-related damping forces are much weaker in comparison to  
those of radiative decay \cite{Mikhailov04} and thus negligible,
whence ${\bm v}_s$ and ${\bm v}_c$ are in fact directly given from Eqs.\,(\ref{2deqv1}) and (\ref{2deqv2})
by the high-frequency electric field  ${\bm E}(t)$ inside the 2D electron system.
On the other hand, by solving the Maxwell equations connecting both sides of the 2D
electron gas which is carrying the sheet current density $N_{s}e{\bm v}(t)$, ${\bm E}(t)$ is determined by the incident fields 
${\bm E}_{{\rm i}s}$ and ${\bm E}_{{\rm i}c}$ based on the setup of the 2D system 
in the sample substrate.\cite{Lei03}

If the 2D electron gas locates within a thin layer 
under the surface plane at $z=0$ of a thick (treated as semi-infinite) 
semiconductor substrate having a refractive index, $n_{s}$, 
the ac field ${\bm E}(t)$ driving the 2D electrons, which equals the sum of the incident
and the reflected fields at $z=0$ and equals the transmitted field (the field 
just passes through the 2D layer), can be expressed as \cite{Chiu-1976} 
\begin{equation}
{\bm E}(t)=\frac{N_{s}e\,{\bm v}(t)}{(n_0+n_{s})\epsilon_0 c}+
\frac{2n_0}{n_0+n_{s}}{\bm E}_{\rm i}(t).\label{thick} 
\end{equation}
Here $n_0$ is the refractive index of
the air and $c$ and $\epsilon_0$ are, respectively, the light
speed and the dielectric constant in vacuum.  
If the 2D electron gas
is contained in a thin layer suspended in vacuum at the plane $z=0$,
then
\begin{equation}
{\bm E}(t)=\frac{N_{s}e\,{\bm v}(t)}{2\epsilon_0 c}+
{\bm E}_{\rm i}(t).\label{thin}
\end{equation}
By combining Eq.\,(\ref{thick}) or Eq.\,(\ref{thin}) with balance Eqs.\,(\ref{2deqv1}) and (\ref{2deqv2}), 
the ac drift-velocity components ${\bm v}_s$ and ${\bm v}_c$, and thus the argument $\xi$, can be determined 
by the incident field components ${\bm E}_{{\rm i}s}$ and ${\bm E}_{{\rm i}c}$. 
 
The imaginary part of 2D electron density correlation function in a magnetic field, 
${\it \Pi}_2({q}_{\|}, {\it \Omega})$, can be written in the Landau representation as \cite{Ting}
\begin{eqnarray}
&&\hspace{-0.7cm}{\it \Pi}_2({q}_{\|},{\it \Omega}) =  \frac 1{2\pi
l_{B}^2}\sum_{n,n'}C_{n,n'}(l_{B}^2q_{\|}^2/2) 
{\it \Pi}_2(n,n',{\it \Omega}),
\label{pi_2q}\\
&&\hspace{-0.7cm}{\it \Pi}_2(n,n',{\it \Omega})=-\frac2\pi \int d\varepsilon
\left [ f(\varepsilon )- f(\varepsilon +{\it \Omega})\right ]\nonumber\\
&&\,\hspace{2cm}\times\,\,{\rm Im}G_n(\varepsilon +{\it \Omega})\,{\rm Im}G_{n'}(\varepsilon ),
\label{pi_2ll}
\end{eqnarray}
where $l_{B}=\sqrt{1/|eB|}$ is the magnetic length,
$
C_{n,n+l}(Y)\equiv n![(n+l)!]^{-1}Y^l{\rm e}^{-Y}[L_n^l(Y)]^2
$
with $L_n^l(Y)$ being the associate Laguerre polynomial, $f(\varepsilon
)=\{\exp [(\varepsilon -\mu)/T_{e}]+1\}^{-1}$ is the Fermi 
function at electron temperature $T_{e}$, 
and ${\rm Im}G_n(\varepsilon )$ is the density-of-states (DOS) function of the broadened Landau level $n$.

The Landau-level broadening, which results from impurity, phonon, and electron-electron scatterings,
is assumed to have a Gaussian form [$\varepsilon_n=(n+\frac{1}{2})\omega_c$ 
is the center of the $n$th Landau level, $n=0,1,2,...$],
\begin{equation}
{\rm Im}G_n(\varepsilon)=-(2\pi)^{\frac{1}{2}}{\it \Gamma}^{-1}
\exp[-2(\varepsilon-\varepsilon_n)^2/{\it \Gamma}^2],
\label{gauss}
\end{equation}
with a $B^{\frac{1}{2}}$-dependent half width expressed as
\begin{equation}
{\it \Gamma}=(2\omega_c/\pi \tau_s)^{\frac{1}{2}}, 
\label{gamma12}
\end{equation} 
where $\tau_s$, the single-particle lifetime or quantum scattering time in the zero magnetic field,
is assumed to be related to the transport relaxation time $\tau_{\rm tr}$ or the zero-field linear mobility $\mu_0$
using an empirical parameter $\alpha$ by \cite{Lei03}
\begin{equation}
1/\tau_s=4\alpha/\tau_{\rm tr}=4 \alpha e/m \mu_0. \label{alpha}
\end{equation}

For an isotropic system where the frictional force ${\bm F}$ is in the opposite direction of 
the drift velocity ${\bm v}$
and the magnitudes of both the frictional force and the energy-dissipation rate depend only on 
$v\equiv |{\bm v}|$, we can write ${\bm F}({\bm v})=F(v){\bm v}/v$ and 
$W({\bm v})=W(v)$.  
In the Hall configuration with velocity ${\bm v}$ in the $x$ direction
${\bm v}=(v,0,0)$ or the current density $J_x=J=N_{s}ev$, and $J_y=0$,
the  longitudinal linear ($v\rightarrow 0$) resistivity under the incident radiation of Eq.\,(\ref{inci})
due to impurity scattering,  
can be written in the form  
\begin{equation}
R_{xx}=-\sum_{{\bm q}_\|}q_x^2
\frac{|U(q_\|)|^2}{N_{s}^2 e^2}\!\sum_{n=-\infty }^\infty \!{J}_n^2(\xi)\,
{\it \Pi}_2^{\prime} ( q_{\|},n\omega), \label{rixx}
\end{equation}
where ${\it \Pi}_{2}^{\prime}({q}_\|,{\it \Omega})\equiv \partial {\it \Pi}_{2}({q}_\|,{\it \Omega})/\partial {\it \Omega}$.

\section{Oscillatory resistivity versus polarization direction}

We consider an incident radiation of form (\ref{inci}) linearly polarized along the direction having an angle $\theta$
with respect to the $x$-axis: ${\bm E}_{{\rm i}s}=(E_{{\rm i}\omega}\cos\theta, E_{{\rm i}\omega}\sin\theta)$ and 
${\bm E}_{{\rm i}c}=0$. The 2D electron gas, having carrier sheet density $N_s=2.2\times 10^{15}$\,m$^{-2}$ and 
zero-temperature linear mobility $\mu_0=800$\,m$^{2}$/V\,s 
from short-range impurity scattering in the absence of the magnetic field, is assumed to locate within a thin layer 
under the surface plane at $z=0$ of a thick GaAs-based substrate with a refractive index of $n_{s}=3.59$.

\begin{figure}
\includegraphics [width=0.38\textwidth,clip=on] {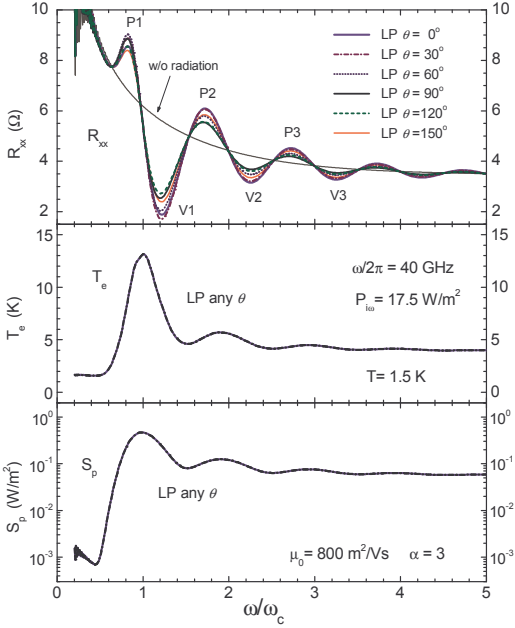}
\vspace*{-0.2cm}
\caption{(Color online) Magnetoresistivity $R_{xx}$, electron temperature $T_{e}$ and energy absorption $S_{\rm p}$
are plotted versus $\omega/\omega_c$
under the irradiation of linearly polarized (LP) microwaves of frequency $40$-GHz and incident 
power $P_{{\rm i}\omega}=17.5$\,W/m$^2$ at several different polarization directions 
($\theta=0^{\rm o}, 30^{\rm o}, 60^{\rm o}, 90^{\rm o}, 120^{\rm o}$ and $150^{\rm o}$)
for a system described in the text at temperature $T=1.5$\,K.}
\label{fig1}
\end{figure}

\begin{figure}
\includegraphics [width=0.48\textwidth,clip=on] {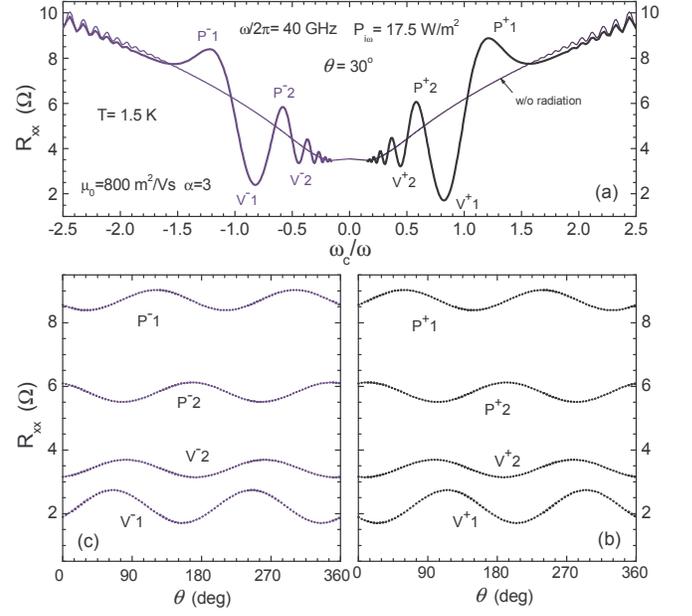}
\vspace*{-0.2cm}
\caption{(Color online) Magnetoresistivity $R_{xx}$ induced by linearly polarized microwave of frequency $40$\,GHz and incident power
$P_{{\rm i}\omega}=17.5$\,W/m$^2$ at $\theta=30^{\rm o}$ is shown for positive and reverse magnetic fields over the
range $-2.5\leq \omega_{c}/\omega \leq 2.5$ (a). 
The $\theta$ dependence of $R_{xx}$ at extrema $P^{+}1$, $V^{+}1$, $P^{+}2$, and $V^{+}2$ (b)
and at extrema $P^{-}1$, $V^{-}1$, $P^{-}2$, and $V^{-}2$ (c).}
\label{fig2}
\end{figure}

Figure 1 shows the calculated energy absorption rate $S_{\rm p}$, the electron temperature $T_{e}$ and
the longitudinal linear resistivity $R_{xx}$ of this system irradiated by linearly polarized (LP) incident microwaves 
having frequency $\omega/2\pi=40$\,GHz and amplitude $E_{{\rm i}\omega}=1.15$\,V/cm   
(i.e., incident power $P_{{\rm i}\omega}=17.5$\,W/m$^2$) at different polarization directions 
($\theta=0^{\rm o}, 30^{\rm o}, 60^{\rm o}, 90^{\rm o}, 120^{\rm o}$, and $150^{\rm o}$), together with
the dark resistivity,
as functions of the normalized inverse magnetic field 
$\omega/\omega_c$ ($\omega_c=eB/m$). The lattice temperature is assumed to be $T=1.5$\,K
and the Landau level broadening parameter is taken to be $\alpha=3$.

We see that the electron energy absorption $S_{\rm p}$ and thus the electron temperature $T_{e}$,
both showing a marked main peak at cyclotron resonance $\omega/\omega_c = 1$ and secondary peaks 
around its harmonics $\omega/\omega_c = 2, 3, 4, ...$, are essentially the same for all different polarization directions of radiation. 
This can be understood directly from expression (\ref{exsp}) of $S_{\rm p}$ in the case of relatively low strength of incident microwave
when the dominant contribution to it comes from the terms $n=\pm 1$ and $J_{\pm 1}^{2}(\xi ) \sim \xi^{2}$.
Writing explicitly the $\theta$-dependent expression of $\xi^{2}$ 
in the case of incident plane-polarized radiation having an angle $\theta$ with respect to the $x$-axis
we can see that, after summing over all the directions of ${\bm q}_\|$, the angle dependence
disappears due to $\sin^2\theta+\cos^2\theta = 1$. Furthermore, the energy-loss rate $W$ has 
$\theta$-dependent behavior similar to that $S_{\rm p}$. Thus
the electron temperature $T_{e}$, determined by the energy-balance equation (7) 
involving only $S_{\rm p}$ and $W$, has similar $\theta$-dependent behavior.
The situation is different for the frictional force ${\bm F}$ and the resistivity $R_{xx}$, because of additional
${\bm q}_\|$-direction dependent weighted factors showing up inside the ${\bm q_\|}$ summation in the expressions (\ref{exf0})
and (\ref{rixx}), leading to sensitive $\theta$-dependence of $R_{xx}$ as seen from the numerical results shown in the upper part of Fig.\,1.    
The linear-polarized microwave excited magnetoresistivity $R_{xx}$ oscillates strongly, having 
the same oscillatory period and nearly the same phase and, at cyclotron resonance $\omega/\omega_c = 1$ 
or its harmonics $\omega/\omega_c =2,3,4,...$,  taking equal values  for all different polarization directions $\theta$.
We can label, for all $\theta$, the main peaks and valleys of the oscillatory resistivity respectively as $P1$, $P2$, and $V1$, $V2$, etc.
The amplitude of resistivity oscillation, however,
varies sensitively with changing the polarization direction of the microwave. 
The maximal or minimal amplitude of a peak or a valley 
shows up at different polarization angles $\theta$ for different peaks or valleys.
For instance, the maximal and minimal amplitudes show up, respectively, at $\theta\approx 59^{\rm o}$ and $\theta\approx 149^{\rm o}$ 
for $P1$, at  $\theta\approx 25^{\rm o}$ and $\theta\approx 115^{\rm o}$ 
for $V1$, at  $\theta\approx 12^{\rm o}$ and $\theta\approx 102^{\rm o}$ 
for $P2$, at  $\theta\approx 8^{\rm o}$ and $\theta\approx 98^{\rm o}$ 
for $V2$, and show up at polarization angles closer to $\theta= 0^{\rm o}$ 
and $\theta= 90^{\rm o}$ for higher order peaks and valleys.
The exact polarization angle for the maximal or minimal amplitude of an extremum 
appears to depend on the radiation frequency and the 2D electron setup in the substrate.
In obtaining the above results the sample setup is so assumed that
2D electrons locate within a thin layer 
under the surface plane at $z=0$ of a thick  
semiconductor substrate having a refractive index of $n_{s}=3.59$, and  
the relevant damping in question, i.e., the radiative decay, is fully determined by this sample setup
from the electrodynamic equation (\ref{thick}). Except for this, no other damping parameter 
nor any mechanism capable of producing polarization rotation is introduced in the present model.

 Next we present the results obtained under magnetic field reversal. 
 Figure 2(a) plots the longitudinal resistivity $R_{xx}$ vs $\omega_{c}/\omega$  over the range
 $-2.5\leq \omega_{c}/\omega \leq 2.5$ for the system irradiated by a linearly polarized microwave along  
 the $\theta=30^{\rm o}$ direction with frequency $\omega/2\pi=40$\,GHz and incident power $P_{{\rm i}\omega}=17.5$\,W/m$^2$.
 The extrema of interest here are labeled as $P^{+}1$, $V^{+}1$, $P^{+}2$, and $V^{+}2$ for those in the domain $B>0$,
 and $P^{-}1$, $V^{-}1$, $P^{-}2$, and $V^{-}2$ for those in the domain $B<0$. The  values of longitudinal resistivity $R_{xx}$
 at $P^{+}1$, $V^{+}1$, $P^{+}2$, and $V^{+}2$ are plotted in Fig.\,2(b), and those at $P^{-}1$, $V^{-}1$, $P^{-}2$, and $V^{-}2$ 
 qre plotted in Fig.\,2(c), as functions of the microwave polarization direction $\theta$.
 The present treatment is for an isotropic system. By symmetry we always have $P^{+}1(\theta)=P^{+}1(\pi+\theta)$, 
 $P^{-}1(\theta)=P^{-}1(\pi+\theta)$, $P^{+}1(\theta)=P^{-}1(\pi-\theta)$, etc. 
 If the maximal amplitude of an extremum, e.g., $P^{+}1$, shows up at $\theta \neq 0$, the 
 maximal amplitude of the corresponding extremum at the reverse magnetic field, $P^{-}1$, must be at a different polarization angle.
 The effect of asymmetry in a real sample itself would produce further complexity.

\section{Summary}

We have examined the effect of changing the polarization angle $\theta$ of the incident linearly-polarized microwaves 
with respect to the dc current on radiation-induced magnetoresistance oscillations 
within the balance-equation formulation of the photon-assisted magnetotransport model.

The present investigation takes the incident microwave field, rather than 
the ac field inside the 2D electron system, as the input quantity, allowing  
a direct determination of the dominant damping mechanism in the high-mobility system, the radiative decay, 
from the experimental sample setup without introducing any artificial damping parameter,
thus enabling a direct comparison with experimental measurement. 

We find that the amplitude of radiation-induced magnetoresistance oscillation
varies sensitively with changing $\theta$.  
At an extremum the amplitude of oscillatory magnetoresistance $R_{xx}$ exhibits a sinusoidal, 
up to a factor of 5, magnitude variation with rotating the polarization angle $\theta$.
The maximal amplitude shows up generally at a nonzero $\theta$, which is dependent upon the extremum in
question, the 2D electron setup, the radiation frequency and the magnetic field orientation.

\vspace{0.1cm}

This work was supported by the National Science Foundation of China (Grant No.~61131006)
and the National Basic Research Program of China (Grant No.~2011CB925603).

\end{document}